\documentclass[11pt]{article}

\usepackage{amsbsy,anysize,comment,amsmath,amssymb}
\marginsize{1.5in}{1.5in}{1.25in}{1.25in}

\def\rmd{{\rm d}}

\begin{document}

\title{\Large\bf An exact solution to a $\boldsymbol{k}$th order, $\boldsymbol{n}$th degree nonlinear
differential equation}

\author{\bf C. Radhakrishnan Nair\\
Centre for Mathematical Sciences,\\
Thiruvananthapuram,\\
PIN - 695 581, INDIA}

\date{}

\maketitle

\baselineskip18pt	

\begin{abstract}
An explicit analytic solution to the nonlinear differential equation
$\displaystyle\left(\frac{\rmd^ky}{\rmd x^k}\right)^n=y^l$ is obtained for arbitrary integer values of $k,l$ and $n$.
\end{abstract}

\section*{Introduction}

Quantum mechanics, relativity and nonlinear physics are often
pointed out~\cite{1} as the landmarks of the 20th century
physics. The development of nonlinear physics occurred mostly in
the last quarter of the 20th century. Chaos, Solitons and
Fractals are the new disciplines that emerged out of the
nonlinear physics. Industrial and technological exploits of the
nonlinear phenomena are now common. Nonlinear differential
equations is one of the tools for the study of nonlinear
phenomena~\cite{2,3}. Therefore, naturally more and more
nonlinear differential equations are set up and methods of
solving nonlinear differential equations are being vigorously
pursued.

It is often very difficult to find exact solutions of nonlinear
differential equations. The higher the nonlinearity, the greater
the difficulty in getting an exact solution. Again, the higher
the order of the nonlinear differential equation, the process of
solving the nonlinear differential equation gets still more
complicated. Owing to the lack of knowledge of solving the
nonlinear differential equations exactly, more and more people
try numerical methods to solve nonlinear differential equations.
But no numerical solution, however great the accuracy, is a
substitute for the analytical solutions. Therefore, even a small
contribution to the analytical solution of differential equation
is a welcome step.

\section*{The Nonlinear Differential Equation and its Solution}

We consider the following nonlinear differential equation of the
$k$th order and $n$th degree.
\begin{equation}\label{eqn1}
\left(\frac{\rmd^ky}{\rmd x^k}\right)^n=y^l
\end{equation}
where $k,l,n$ are integers and $l\neq n$.

Let us assume that a solution of the form 
\begin{equation}\label{eqn2}
y=ax^m
\end{equation}
exists, then
$$
\frac{\rmd^ky}{\rmd x^k}=[am(m-1)(m-2)\dots\{m-(k-1)\}]x^{m-k}
$$
and
$$
y^l=a^lx^{ml}
$$
substituting the values of $\frac{\rmd^ky}{\rmd x^k}$ and $y^l$
into equation~(\ref{eqn1}), we get
\begin{equation}\label{eqn3}
[\{am(m-1)(m-2)\dots(m-(k-1)\}x^{m-k}]^n=a^lx^{ml}
\end{equation}

Equating the powers of $x$, on the left hand side and right hand
side of equation~(\ref{eqn3}), we get
\begin{eqnarray}
&&n(m-k)=ml\nonumber\\
\therefore&&m=\frac{nk}{n-l}\label{eqn4}
\end{eqnarray}
Equating the coefficients of $x$,
\begin{eqnarray}
a^nm^n(m-1)^n(m-2)^n\dots\{m-(k-1)\}^n=a^l\nonumber\\
a=[m(m-1)(m-2)\dots(m-(k-1)\}]^{\frac{n}{l-n}}\label{eqn5}
\end{eqnarray}
Since $a$ and $m$ are known the solution
$$
y=ax^m\mbox{ is known}
$$
Illustration

By putting $k=2$, $l=4$ and $n=6$ in (\ref{eqn1}), we get the
equation 
\begin{equation}\label{eqn6}
\left(\frac{\rmd^ky}{\rmd x^k}\right)^6=y^4
\end{equation}
From equation (\ref{eqn5}) $a=[6^6\cdot 5^6]^{-1/2}$.

Substituting the values of $m$ and $a$ in the solution
(\ref{eqn2}), we get
\begin{equation}\label{eqn7}
y=\frac{1}{\sqrt{6^65^6}}x^6
\end{equation}
It is verified that equation~(\ref{eqn7}) is a solution of
equation~(\ref{eqn6}).

\section*{Discussion}

The solution we found for equation~(\ref{eqn1}) is simple. But
the importance of this solution is that we can find the solution
for any order and any degree of the differential equation. This
implies that a very large class of physical phenomena can be
studied using equation~(\ref{eqn1}). If we know one of the
solutions of a nonlinear differential equation, this knowledge
may prove useful to generate some other solutions.

\section*{Acknowledgement}

The author wishes to express his gratitude towards E.~Krishnan
and V.~Vipindas for fruitful discussions and to L.~A.~Ajith for
beautifully putting the material in \LaTeX. General support from
Focal Image (India) Pvt Ltd was always a source of strength.

\end{document}